\documentclass[prc,aps,showpacs,nofootinbib,twocolumn,preprintnumbers]{revtex4}
\usepackage{latexsym}
\usepackage{amsmath, mathrsfs}
\usepackage[latin1]{inputenc}
\usepackage[dvips]{graphicx}

\newcommand{\qsa}{{Q^A_\mathrm{s}}}
\newcommand{\qsb}{{Q^B_\mathrm{s}}}
\newcommand{\qs}{{Q_\mathrm{s}}}
\newcommand{\qss}{{Q_\mathrm{s}^2}}
\newcommand{\qsc}{{Q_\mathrm{s}^3}}

\newcommand{\xt}{\mathbf{x}_T}
\newcommand{\bt}{\mathbf{b}_T}

\newcommand{\pt}{{\mathbf{p}_T}}

\newcommand{\ra}{{R_A}}
\newcommand{\gev}{\ \textrm{GeV}}
\newcommand{\fm}{\ \textrm{fm}}

\newcommand{\npart}{{N_\textrm{part}}}
\newcommand{\nparta}{{N^A_\textrm{part}}}

\newcommand{\ncoll}{{N_\textrm{coll}}}

\newcommand{\nr}[1]{(\ref{#1})} 
\newcommand{\ud}{\, \mathrm{d}}
\newcommand{\fig}{Fig.~}
\newcommand{\eq}{Eq.~}

\newcommand{\eqs}{Eqs.~}

\begin{document}

\title{
Universality of the saturation scale and the initial eccentricity in heavy ion collisions
}

\preprint{nucl-th/0609021}

\author{T. Lappi}
\affiliation{Physics Department, Brookhaven
National Laboratory, Upton, NY 11973, USA}
\author{R. Venugopalan}
\affiliation{Physics Department, Brookhaven
National Laboratory, Upton, NY 11973, USA}

\begin{abstract} 
Recent estimates that Color Glass Condensate initial conditions may generate  a
larger initial eccentricity for noncentral relativistic heavy ion
collisions (relative to the initial eccentricity assumed in earlier 
hydrodynamic calculations) have raised the possibility of a higher bound on the
viscosity of the Quark Gluon Plasma. We show that this large initial eccentricity results in part from a definition of the saturation scale as
proportional to the number of nucleons participating in the collision. A 
saturation scale proportional to the nuclear thickness function (and therefore 
independent of the probe) leads to a smaller eccentricity, albeit still larger 
than the value used in hydrodynamic models. Our results suggest that the early 
elliptic flow in heavy ion collisions (unlike multiplicity distributions) is sensitive to the universality of the saturation scale in high energy QCD.

\end{abstract}


\maketitle

\section{Introduction}

Data from RHIC on bulk 
properties~\cite{Adcox:2004mh,Adams:2005dq,Arsene:2004fa,Back:2004je} of 
the matter produced in heavy ion collisions, is qualitatively described 
by hydrodynamic calculations~\cite{Huovinen:2006jp}. Increasingly 
precise experimental data now enables us to probe details of initial 
conditions for the hydrodynamical calculations.  These initial 
conditions may be obtained from the Color Glass Condensate (CGC) 
framework~\cite{Iancu:2003xm} describing the phenomenon of parton saturation~\cite{Gribov:1984tu,Mueller:1985wy} 
in the high energy nuclear wavefunction.

One particular application of the CGC framework 
to the early stages of heavy ion collisions has been the formulation of 
the problem as a solution of classical field 
equations~\cite{Kovner:1995ts,Kovner:1995ja, Kovchegov:1997ke, 
Gyulassy:1997vt,Dumitru:2001ux} 
and the numerical solution of these equations \cite{% 
Krasnitz:1998ns,Krasnitz:1999wc,Krasnitz:2000gz,Krasnitz:2001qu,Krasnitz:2003jw,% 
Krasnitz:2002ng,Krasnitz:2002mn,Lappi:2003bi,Lappi:2004sf,
Lappi:2006hq,Romatschke:2005pm,Romatschke:2006nk}. By solving 
classical equations, one efficiently sums all leading order Feynman 
graphs contributing to the inclusive multiplicity. We shall henceforth 
refer to this approach as the Classical Yang-Mills (CYM) approach.  
An alternative approach is 
to assume the framework of $k_\perp$ factorization, where the inclusive 
multiplicity of produced gluons is expressed as a convolution of 
unintegrated gluon distributions from the projectile and the target. 
While exact to leading order for proton-nucleus 
collisions~\cite{Braun:2000bi,Kharzeev:2003wz,Blaizot:2004wu}, this 
approach is only approximate in heavy ion collisions at central 
rapidities.  A popular version of the $k_\perp$ factorization approach 
applied to heavy ion collisions, with a particular ansatz for the 
unintegrated gluon distributions, is the KLN 
approach~\cite{Kharzeev:2001gp,Kharzeev:2002pc,Kharzeev:2002ei,% 
Kharzeev:2000ph,Kharzeev:2004if}. The solution of the CYM equations incorporates, to lowest order in the coupling, 
all terms that violate $k_\perp$ factorization; these solutions however are numerically intensive. On the other hand, the KLN 
ansatz, while approximate, provides an analytic expression that captures some key features of saturation. 

A striking signal of collective hydrodynamical behavior of the matter 
produced at RHIC is elliptic 
flow~\cite{Adams:2004bi,Adler:2003kt,Back:2004mh,Ito:2005rt}. The large 
elliptic flow observed at RHIC requires strong interactions during the 
first fermis of the collision.  The elliptic flow in this strongly 
interacting system is particularly sensitive to the initial eccentricity 
(the anisotropy of the energy density in the transverse plane), 
\begin{equation}\label{eq:eccdef} \epsilon = \frac{\int \ud^2 \xt\, 
\varepsilon(\xt)\, (y^2 - x^2) }{\int \ud^2 \xt\, \varepsilon(\xt)\, 
(y^2 + x^2) }, \end{equation} where $\varepsilon$ is the local energy 
density~\footnote{Alternately, the eccentricity is defined as the anisotropy of the 
entropy density in some works~\cite{Bhalerao:2005mm}. As we will see, this definition is less sensitive to different CGC initial conditions. It is  
however not the definition used in comparisons of detailed hydrodynamic models to RHIC data.}.
In a hydrodynamical description of heavy ion collisions, the 
eccentricity of the initial condition is the primary factor determining 
the elliptic flow observed in the final state \cite{Ollitrault:1992bk}.

In this short note, we will study how the eccentricity is computed in 
the CGC framework in the CYM and KLN models.  We shall first discuss 
different ways used in the literature to understand the transverse 
coordinate dependence of the saturation scale, in particular based on 
its \emph{universality}.  We shall then explain how these lead to 
different results for $\epsilon$, in particular the recent results in 
Refs.~\cite{Hirano:2005xf,Drescher:2006pi}. (See also 
Ref.~\cite{Kuhlman:2006qp} for a discussion of related issues for 
uranium-uranium collisions.)

\section{A universal saturation scale}

In the CGC framework, the production of particles in the initial stage 
of a heavy ion collision is controlled by one parameter, the saturation 
scale $\qs$. The initial spatial anisotropy of the energy density must 
thus come from the transverse coordinate dependence of the saturation 
scale. In the original MV model 
\cite{McLerran:1994ni,McLerran:1994ka,McLerran:1994vd} the large $x$ 
degrees of freedom are thought of as independent static 
(independent of light cone time) color 
charges with a transverse color charge density $g^2 \mu^2$. The density 
of color charges is a property of the nuclear wavefunction and 
independent of the probe --- it is universal~\cite{Iancu:2001md}.  If 
the color charge density arises as a superposition of independent large 
$x$ partons, it is natural that the density should be proportional to 
the nuclear thickness function $T_A$, thereby providing an average 
measure of the number of these fast partons. It was later shown 
\cite{Jalilian-Marian:1997xn,% 
Kovchegov:1996ty,Kovchegov:1998bi,McLerran:1998nk,Gelis:2001da,Blaizot:2004wu,% 
Blaizot:2004wv} that the MV model does indeed exhibit saturation, with 
$\qs \sim g^2 \mu$, and thus also the saturation scale $\qs$ should be 
universal with 
\begin{equation}\label{eq:qsta} 
\qss (\xt) = \qs_0^2 \frac{\pi \ra^2}{A} T_A(\xt)\, . 
\end{equation} 
Here $\qs_0$ is the average saturation scale, 
$T_A(\xt) = \int_{-\infty}^{\infty} \ud z \,\rho({\bf r})$ 
is the thickness function,  $\xt$ is the coordinate 
relative to the center of the nucleus and $\rho({\bf r})$ is the 
Woods-Saxon nuclear density profile, normalized as 
$\int \ud^3 \mathbf{r}\, \rho(\mathbf{r}) = A$. It has also been argued 
\cite{Mueller:2003bz} that at sufficiently high energy the dependence of 
the saturation scale on $T_A$ and thus $A$ would change due to high 
energy evolution, but we will not consider this possibility here.

The classical field picture of the MV model, with this universally 
defined saturation scale, was applied in both analytical 
\cite{Kovner:1995ts,Kovner:1995ja,Gyulassy:1997vt,Kovchegov:1997ke,Dumitru:2001ux} 
and numerical classical Yang-Mills (CYM)~\cite{% 
Krasnitz:1998ns,Krasnitz:1999wc,Krasnitz:2000gz,Krasnitz:2001qu,%
Krasnitz:2002ng,Krasnitz:2003jw,% 
Krasnitz:2002mn,Lappi:2003bi,Lappi:2004sf,Lappi:2006hq,Romatschke:2005pm,Romatschke:2006nk} computations of 
the ``Glasma'' \cite{Lappi:2006fp} fields in the initial stages of the 
collision, Noncentral collisions with finite impact parameter were 
studied in Refs.~\cite{Krasnitz:2002ng,Lappi:2003bi} and particularly in 
the detailed study of Ref.~\cite{Krasnitz:2002mn}.

In the ``KLN'' approach, 
\cite{Kharzeev:2001gp,Kharzeev:2002pc,Kharzeev:2002ei,Kharzeev:2000ph,Kharzeev:2004if}, 
particle production is computed using unintegrated gluon distribution 
functions depending on the saturation scale $\qs(x,\xt)$. These 
calculations used a saturation scale dependent on the number of 
participant nuclei: 
\begin{equation}\label{eq:qsnpart} 
\qsa^2(\xt) \sim 
\nparta(\xt), 
\end{equation} 
with 
\begin{multline} \nparta(\xt) = 
T_A(\xt) \\ \times
 \left(1- \left(1-\sigma_{NN} 
\frac{T_B(\xt-\bt)}{B} \right)^B \right).
\end{multline}
Note that as this form involves the thickness functions of both nuclei 
$A$ and $B$, it is not manifestly universal. While a nonuniversal saturation
scale has been used in many phenomenological studies, such as
the EKRT ``final state saturation''  model of Ref.~\cite{Eskola:1999fc},
the logic for introducing 
the second term in the original KLN works was not the final state 
picture. The $\npart$ definition was used instead to model the 
interaction probability for the scattering 
in a manner as close to the original Glauber framework as possible. 
There arises however the danger of double counting the interaction 
probability because this probability is already taken into account in 
the derivation of $k_\perp$ 
factorization~\cite{Braun:2000bi,Kharzeev:2003wz,Blaizot:2004wu}.  The 
dangers of overcounting are avoided in the universal prescription for 
$\qs$ in \eq\nr{eq:qsta}.

We must emphasize that the choice between using the $k_\perp$~factorized 
formalism and solving the classical field equations does not dictate how 
the saturation scale depends on the transverse coordinate.  There is no 
technical impediment to using $(\qsa)^2 \sim T_A$ in the $k_\perp$~factorized 
formulation or $(\qsa)^2 \sim \nparta$ in the classical field computation.  
In the numerical study using the classical field equations in 
Ref.~\cite{Krasnitz:2002mn} both prescriptions, \eqs\nr{eq:qsta} 
and~\nr{eq:qsnpart} were considered and were found to give similar 
results for the impact parameter dependence of the multiplicity. 
However, as we shall show in the following, they do yield different 
results for the eccentricity.

\section{How the choice of $\qs$ influences the eccentricity}

\begin{figure} 
\begin{center} 
\includegraphics[width=0.4\textwidth]{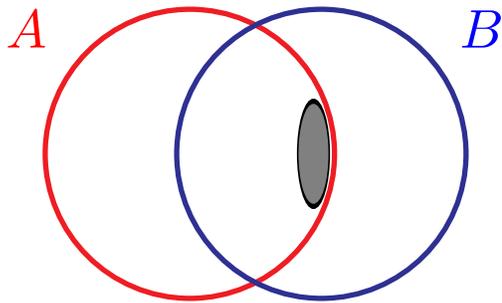} 
\end{center} 
\caption{The area in the transverse plain that is most important for our 
discussion here is marked with the gray oval, near the edge of nucleus 
$A$ and the center of nucleus $B$. The same discussion applies for the 
edge of nucleus $B$ near the center of nucleus $A$.}
\label{fig:nn} 
\end{figure}

For computing the eccentricity, the crucial contributions (in the 
geometry of the transverse plane of the scattering) come from regions 
where the saturation scale in one nucleus is significantly larger than 
in the other. Let us first review what the spectrum of gluons looks like 
in this case, with two saturation scales $\qs_1$ and $\qs_2$ satisfying 
$\qs_1 < \qs_2$ \cite{Dumitru:2001ux,Blaizot:2004wu}. Parametrically, the 
spectrum of produced gluons behaves as 
\begin{eqnarray} 
\frac{\ud N}{\ud^2 \xt \ud^2 \pt} & \sim & \ln(p_T), \quad p_T < \qs_1 
\\ 
& \sim & \frac{\qs_1^2}{p_T^2}, \quad \qs_1 < p_T < \qs_2 
\\ 
& \sim & 
\frac{\qs_1^2 \qs_2^2}{p_T^4}, \quad p_T > \qs_2 . 
\end{eqnarray} 
Integrated over transverse momenta, this gives \begin{eqnarray} 
\label{eq:pamulti} \frac{\ud N}{\ud^2 \xt} &\sim& \qs_1^2 \\ 
\label{eq:paen} \frac{\ud E_T}{\ud^2 \xt} &\sim& \qs_1^2 \qs_2\,, 
\end{eqnarray} neglecting logarithmic corrections $\sim \ln\left( 
\qs_2/\qs_1\right)$. The additional dependence on 
$\qs_2$ in the transverse energy, relative to the multiplicity, holds 
the key to the following discussion.

\begin{figure}
\begin{center}
\includegraphics[width=0.45\textwidth]{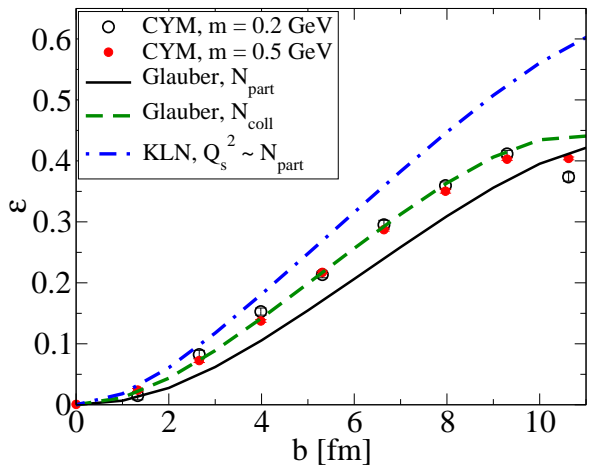}
\end{center}
\caption{The eccentricity as a function of impact parameter. The classical field CGC 
result with two different infrared cutoffs $m$ is denoted by CYM. The 
traditional initial eccentricity used in hydrodynamics is a linear combination
of mostly ``Glauber $\npart$'' and a small amount of 
``Glauber $\ncoll$''. The  ``KLN'' curve is the 
eccentricity obtained from the CGC calculation in 
Refs.~\cite{Hirano:2005xf,Drescher:2006pi}. }
\label{fig:ecc}
\end{figure}

The difference between the two definitions of the transverse coordinate 
dependence of the saturation scale, \eqs\nr{eq:qsta} and~\nr{eq:qsnpart} 
is the largest in the region near the edge of one nucleus (labeled as nucleus $A$) and in the center of the other (nucleus $B$), so 
that $\qsa < \qsb$; the geometry is illustrated in \fig\ref{fig:nn}. The 
smaller saturation scale approaches zero as $\left(\qsa\right)^2 \sim 
T_A$ regardless of the definition of $\qs$ 
(\eq\nr{eq:qsta} or \eq\nr{eq:qsnpart}). But the behavior of the larger 
saturation scale $\qsb$ is different in the two cases. Using the universal definition of 
$\qs$ in 
\eq\nr{eq:qsta} $\qsb$ is large, $(\qsb)^2 \sim T_B$. In contrast, the non-universal 
$\npart$-definition of $Q_s$ in \eq~\nr{eq:qsnpart} suggests that the larger saturation 
scale $\qsb$ 
also approaches zero as $\sigma_{NN} T_A T_B$.

Because the multiplicity \nr{eq:pamulti} only depends on the smaller 
saturation scale $\qsa$, the difference in the gluon multiplicities 
between the two definitions \eqs\nr{eq:qsta} and~\nr{eq:qsnpart} is 
small. This explains the numerical observation in  
Ref.~\cite{Krasnitz:2002mn} that both the KLN prescription for $\qs$ and the universal CYM one 
give very similar results for the centrality dependence of the multiplicity. The larger 
saturation scale $\qsb$ and 
therefore the energy density are, however, very different in the two cases. This 
difference is accentuated in the eccentricity \nr{eq:eccdef}.  With the 
$\npart$ definition~\nr{eq:qsnpart}, the energy density in this edge 
region is suppressed relative to the universal definition in \nr{eq:qsta}, 
thereby leading to a larger eccentricity.

The eccentricities obtained using the  different transverse coordinate 
dependences of the saturation scales are shown in \fig\ref{fig:ecc}.
The CYM eccentricity in the plot is calculated at $\tau=0.25\fm$, while the 
KLN result does not depend on time. 
The KLN $\npart$ definition of $\qs$ leads to the largest eccentricity.  The 
universal CYM definition gives smaller values of $\epsilon$ albeit larger 
than the 
traditional parametrization (used in hydrodynamical model computations) 
where the energy density is taken to be proportional to the number of 
participating nucleons. This result is also shown to be insensitive to two different choices of 
the infrared scale $m$ which regulates the spatial extent of the Coulomb tails of the gluon 
distribution. We observe that the values of $\epsilon$ from 
the CYM computation are close to those obtained from an energy density 
parametrization following binary collisional ($\ncoll$) scaling.  This 
result can be explained qualitatively as follows.  In the classical 
Yang-Mills calculation the total multiplicity of gluons scales as 
$\qss$, where $\qs$ is the dominant transverse momentum scale of the 
produced gluons, depending on both saturation scales $\qsa$ and $\qsb$. 
The multiplicity of produced gluons $\sim \qss$ turns out to be roughly 
proportional to $\npart$.  The energy density, on the other hand, scales 
as $\qsc$, and one expects it to scale as $(\npart)^\gamma$ with some 
$\gamma>1$. It is therefore natural to expect the eccentricity in a 
saturation model to be larger than the traditional one following from 
$\npart$-scaling of the energy density. However, we see no reason in 
general for it to exactly mimic the result from $\ncoll$-scaling.

\begin{figure}
\begin{center}
\includegraphics[width=0.45\textwidth]{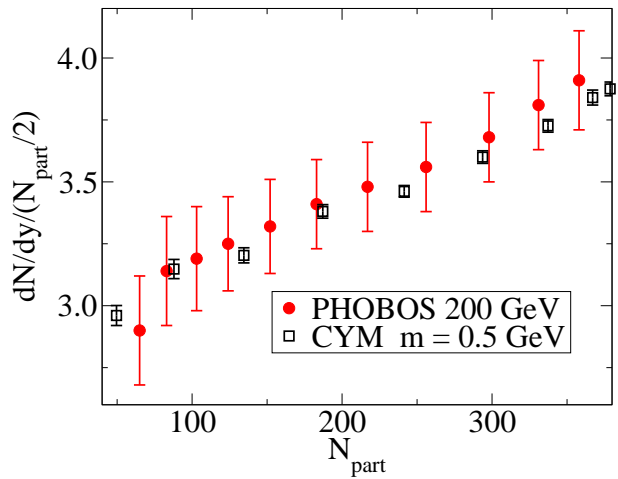}
\end{center}
\caption{The centrality dependence of the multiplicity.
The closed circles are the PHOBOS data for
the charged particle mulciplicity per participant pair
from Ref.~\cite{Back:2002uc}. The open squares are the CYM result for 
gluon multiplicity for $g^2\mu = 1.6 \gev$ and $m=0.5\gev$ and scaled
by $N_\textrm{ch}/N_\textrm{tot} \approx 2/3$. 
 }
\label{fig:mult}
\end{figure}

In \fig~\ref{fig:mult}, we show a plot of the centrality dependence of the multiplicity for 
$g^2\mu =1.6 \gev$ corresponding to an estimated 
gluon multiplicity of $\sim$ 1000 in central Au-Au collisions at 
RHIC~\footnote{The value $g^2\mu=1.6$ 
differs from the previous estimate of 2 GeV~\cite{Krasnitz:2000gz}
primarily because these estimates had very low infrared cut-offs of order 
$m\sim 1/\ra$. For finite
nuclei an infrared scale $m$ of the order of the surface diffuseness of the 
Woods-Saxon density profile is 
required to regulate the Coulomb tails of the gluon field at large distances. 
While the dependence on $m$ is weak, changing it by a factor of 10 does
change the best estimate for $g^2\mu$ by $\sim 20\%$.}.
The  universal $\qss \sim T_A$ 
prescription captures the observed centrality dependence of the multiplicity distribution.  
It has been argued~\cite{AdrianHansYasushi} that 
in a realistic Monte Carlo implementation the KLN formalism can
be recast in a form where the multiplicity is equivalent to one calculated
from universal unintegrated gluon distributions. 
It appears unlikely however that this equivalence holds for other observables.

\section{Conclusions}

We have shown in this brief note that the initial eccentricity of a relativistic heavy ion collision,
computed in the Color Glass Condensate framework, is very sensitive to 
the transverse coordinate dependence of the saturation scale $\qs$. When $\qss$
is proportional to the number of participants $\npart$, the energy density 
produced (near the edge of one nucleus while near the center of the other) is
small, leading to  a large eccentricity. An argument based on the universality
of the saturation  scale leads to $\qss$ in nucleus A being proportional to the nuclear 
thickness function $T_A$ of 
that nucleus alone. In this latter case, the saturation scale in the center of 
one nucleus does not depend on 
the other. While the two definitions have only logarithmic differences for multiplicity 
distributions, the energy density is larger and 
the eccentricity smaller in the universally scaling form of $\qs$ than for the non--universal case.
In both cases, the result for the eccentricity is 
larger than the values  
typically employed in hydrodynamical models which assume ideal hydrodynamics. 
Conceptually, the universality of the saturation scale is important because 
this universality is essential for a reliable framework to compute the 
properties of QCD at high energies.

The larger eccentricities from the CGC initial conditions are of great phenomenological 
interest because they allow for the possibility that there may be significant 
viscous effects in the quark gluon plasma. Our discussion should therefore be taken into
 account in future attempts to place 
an upper bound on the viscosity of the plasma. The reader should note however that these
 eccentricities were computed at very early times 
($\tau\approx 0.25\fm $) after the collision. After this time the eccentricity
will immediately start to decrease because the edges of the system
are expanding into the surrounding vacuum. 
Even if thermalization were to occur at the extremely early times of 
$0.6 \fm$ (as some hydrodynamic models assume), the eccentricity will have decreased from
 the CYM values already at 
$0.25 \fm \rightarrow 0.6 \fm$, which is the time at which one needs the initial conditions.
 Therefore the initial eccentricities being assumed currently 
for hydrodynamic simulations may not too implausible leaving open the possibility that one
 indeed might have a 
``perfect fluid''~\cite{Shuryak:2004cy}. A fuller understanding (and therefore a firmer bound 
on the viscosity) will require a dynamical understanding of how 
eccentricity decreases from the initial ``Glasma'' 
stage to the thermalized stage  of the heavy ion collision. 
\vskip 0.1in
\acknowledgments{We are grateful to Adrian Dumitru, Dima Kharzeev, Genya Levin and 
especially to Hans-Joachim 
Drescher and Yasushi Nara for very valuable and informative discussions.
This manuscript has been authorized under Contract No. DE-AC02-98CH10886 
with the U.S. Department of Energy.}

\bibliographystyle{h-physrev4mod}
\bibliography{spires}

\end{document}